\documentclass[twocolumn,floatfix,aps,amsmath,nofootinbib]{revtex4}
\pdfoutput=1
\usepackage{graphicx}
\usepackage{epstopdf}
\usepackage{amsmath}
\usepackage{amsfonts}
\usepackage{amssymb}
\usepackage{color}
\usepackage{mathrsfs}
\usepackage{multirow}
\usepackage{bm}
\usepackage{bbm} 

\def\half{\frac{1}{2}}
\def\({\left(}
\def\){\right)}
\def\[{\left[}
\def\]{\right]}

\def\e{\begin{equation}}
\def\q{\end{equation}}
\def\m{\begin{eqnarray}}
\def\n{\end{eqnarray}}

\begin{document}

\title{The Large Number Limit of Multifield Inflation}

\author{Zhong-Kai Guo$^{1,2,}$\footnote{guozhk@itp.ac.cn}}
\affiliation{$^1$CAS Key Laboratory of Theoretical Physics, Institute of Theoretical Physics, Chinese Academy of Sciences, Beijing 100190, China\\
$^2$School of Physical Sciences, University of Chinese Academy of Sciences, No. 19A Yuquan Road, Beijing 100049, China}
\date{\today}

\begin{abstract}

We compute the tensor and scalar spectral index $n_t$, $n_s$, the tensor-to-scalar ratio $r$, the consistency relation $n_t/r$ in the general monomial multifield slow-roll inflation models with potentials $V \sim\sum_i\lambda_i \left|\phi_i\right|^{p_i}$. The general models give a novel relation that $n_t$, $n_s$ and $n_t/r$  are all proportional to the logarithm of the number of fields $N_f$ when $N_f$ is getting extremely large with the order of magnitude around $\mathcal{O}(10^{40})$. An upper bound $N_f\lesssim N_*e^{ZN_*}$ is given by requiring the slow variation parameter small enough where $N_*$ is the e-folding number and $Z$ is a function of distributions of $\lambda_i$ and $p_i$. Besides, $n_t/r$ differs from the single-field result $-1/8$ with substantial probability except for a few very special cases. Finally, we derive theoretical bounds $r>2/N_*$ ($r\gtrsim0.03$) and for $n_t$ which can be tested by observation in the near future.

\end{abstract}


\maketitle

\newpage

\section{Introduction}
Besides the quantum fluctuations generated during inflation which seed the anisotropies of cosmic microwave background (CMB) and the large-scale structure (LSS) observed in today's Universe \cite{inflationPerturbation,komatsu_seven-year_2011,Planck:2013jfk,planck_collaboration_planck_2015-4}, another significant prediction of inflation models is the primordial gravitational wave. Although astrophysical gravitational waves produced by binary black holes have been detected by LIGO \cite{LIGO}, the searching for primordial gravitational wave and its contribution to CMB B-mode polarization is still underway at present and in the future \cite{planck2015,Remazeilles:2017szm,Matsumura:2013aja, Baumann:2008aq, Andre:2013afa, 
BICEP2-2016, Grayson:2016smb, Henderson:2015nzj, Harrington:2016jrz, Benson:2014qhw, Arnold:2014:Simons, Aboobaker:2017EBEX, Nagy:2017csq, Masui:2010cz, Book:2011dz}.

The consistency relation \cite{consistencyRelation} in single-field slow-roll inflation $n_t/r=-1/8$ is a relation between tensor spectral index $n_t$ and the tensor-to-scalar ratio $r$. It is hoped that the detection of such a compelling signature can further validate inflation theory especially for the single-field ones. Unfortunately the excess of B-mode power detected by BICEP2 \cite{Ade:2014xna} can be explained by the polarized thermal dust, not the primordial gravitational wave \cite{Mortonson:2014bja,Flauger:2014qra,Cheng:2014pxa}. 

Recent experimental progress however, especially the \textit{Planck}~2015 results \cite{planck2015} shows $r_{0.002}<0.11$ at $95\%$ C.L. by fitting the \textit{Planck} TT,TE,EE+lowP+lensing combination (P15). The BICEP2 \& \textit{Keck Array} B-mode data (BK14) implies $r_{0.05}<0.09$ ($95\%$ C.L.). Combining with \textit{Planck}~2015 TT+lowP+lensing and some other external data, the upper bound on $r$ becomes $r_{0.05}<0.07$ ($95\%$ C.L.) \cite{BICEP2-2016,Huang:2015cke} in the base $\Lambda$CDM+$r$ model. The tight constraint on $r$ lead to the chaotic single-field inflation model with a potential $V(\phi)\propto \phi^2$ being disfavored at more than $2\sigma$ C.L. \cite{Cheng:2014pxa}. Moreover, single-field inflation models with a monomial potential and the natural inflation model are all marginally disfavored at 95\% C.L.  and all single-field inflation models with a convex potential are not favored \cite{Huang:2015cke}.

On the other hand, many high energy theories contain large numbers of scalar degrees of freedom in extremely high energy scales \citep{Grana:2005jc,Douglas:2006es,Denef:2007pq,Denef:2008wq}, therefore single-field inflation models are simple but not natural in the very early Universe in approaching Planck energy density. Consequently, the studies of gravitational wave consistency relation and other inflationary observables for large-number multifield inflation are necessary and may provide a better representation of our real Universe. Price {\it et al.} derived good results for the $N_f$-monomial models with potential $V\sim\sum_i\lambda_i|\phi_i|^p$ by marginalizing probability random method and many-field limit \cite{Price:2014ufa}. However, diverse exponent $p$ will be more appropriate and more consistent with many high energy theories \cite{Liddle:1998jc,Kanti:1999vt,Kanti:1999ie,Kaloper:1999gm,Easther:2005zr,Dimopoulos:2005ac,Kim:2006ys,Kim:2007bc}.

In this paper we will derive robust results for $n_t/r$ and other inflationary parameters in $N_f$-monomial models with diverse exponent $p_i$. In Sec. \ref{basics}, we employ $\delta$N-formalism, central limit theorem (CLT) and Laplace method sequentially to calculate the expectations and corresponding variances of all the inflation parameters and prove their robustness. Numerical verifications and intuitive graphic representations are showed for some well-motivated prior probabilities of $\lambda_i$, $p_i$ and initial conditions. We conclude in Sec. \ref{conclusions}. 

\section{The General Large-$N_f$ Monomial multifield  Models}\label{basics}
We consider the multifield inflation with potential
\begin{align}\label{potential}
V=\sum_{i=1}^{N_f} V_i(\phi_i)=\sum_{i=1}^{N_f} \lambda_i |\phi_i|^{p_i},
\end{align}
where $\phi_i$ is the inflaton field, $N_f$ is the number of fields 
and $\lambda_i$, $p_i$ are real, positive constants. For simplicity, we set the reduced Planck mass $M_{pl}=1/\sqrt{8\pi G}\equiv 1$.

According to the slow-roll inflation of first-order approximation we have $n_t=-2\epsilon$ and
\begin{align}
\epsilon=\half \sum_i \left(\frac{V_i'}{V}\right)^2,
\end{align}
where $V'_i\equiv dV_i/d\phi_i$.

In $\delta$N-formalism, applying initial flat slice of space-time at time $t_*$ gives the number of e-folds $N_*$ from $t_*$ when the pivot scale $k_*$ leaves the horizon to the end of inflation at $t_c$ \cite{Vernizzi:2006ve,Battefeld:2006sz} 
\begin{align}
N_*=-\int_{\phi_*}^{\phi_c}\sum_i^{N_f}\frac{ V_i}{V_i'}d\phi_i,
\end{align}
where $\phi_{i,*}$ and $\phi_{i,c}$ denote values at the horizon crossing time and the end of inflation respectively. Substitute $V_i$ and $V_i'=\lambda_i p_i|\phi_i|^{p_i-1}$ then
\begin{align}\label{NtoField}
N_*=\half\sum_i^{N_f}\frac{1}{p_i}(\phi_{i,*}^2-\phi_{i,c}^2).
\end{align}

We can also express the gauge-invariant curvature perturbation $\zeta$ by the field perturbations at horizon crossing
$\zeta\approx\sum_iN_{*,i}\delta\phi_{i,*}$, 
where $N_{*,i}\equiv \partial N_{*}/\partial\phi_{i}$. The power spectrum of scalar field perturbations around a smooth background at time $t_*$ is 
$P^{ij}_{\delta\phi}=\left(H_*/2\pi\right)^2\delta^{ij}$.
Consequently, the power spectrum of curvature perturbation is
\begin{align}
P_{\zeta}=\sum\nolimits_iN_{*,i}N_{*,i}\left(\frac{H_*}{2\pi}\right)^2.
\end{align}
Recalling the tensor power spectrum $P_{h}=2H^2_*/\pi^2$ finally comes to the expression of tensor-to-scalar ratio in $\delta$N-formalism
\begin{align}
r=\frac{8}{\sum\nolimits_iN_{*,i}N_{*,i}}.
\end{align}

For $N_f$-monomial models, it is reasonable to neglect the field values $\phi_{i,c}$ at the end of inflation, i.e., we apply the horizon crossing approximation (HCA). 
From the definitions of scalar spectral index $n_s$ and $\epsilon$ we can derive $ n_s-1 =d\ln \sum\nolimits_i N_{*,i}N_{*,i}/dN-2\epsilon$,
where we have taken the first order approximation of $\epsilon$. Substituting the Friedman equations, the Klein-Gorden equations and the relation $dN=Hdt$ comes to
\begin{align}
               \frac{d\ln \sum\nolimits_i N_{*,i}N_{*,i}}{dN}=\frac{-2}{\sum\nolimits_i N_{*,i}N_{*,i}}\frac{1}{V}\sum\nolimits_i V_i'N_{*,i}N_{*,ii}\nonumber.
 \end{align}
In our general $N_f$-monomial model the $V$ and $N_*$ are Eq. \eqref{potential} and Eq. \eqref{NtoField} respectively and using HCA then gives the scalar spectral index $n_s$ in the first order of $\epsilon$
\begin{align}
n_s-1=-\frac{2}{\sum\nolimits_i\phi_{i}^2/p_i^2}\frac{\sum\nolimits_j(\lambda_j/p_j)|\phi_j|^{p_j}}{\sum\nolimits_{l} \lambda_l |\phi_l|^{p_l}}-2\epsilon,\label{ns}
\end{align}
where $\phi_i$ means the field values $\phi_{i,*}$ at horizon crossing.
Other inflation parameters in explicit expressions of $\phi_i$, $p_i$, $\lambda_i$ are as follows:
\begin{align}
\epsilon&=\frac{1}{2} \frac{\sum\nolimits_i\lambda_i^2 p_i^2|\phi_i|^{2p_i-2}}{\left(\sum\nolimits_{j} \lambda_j |\phi_j|^{p_j}\right)^2},\label{slow-roll}\\
     n_t&=-\frac{\sum\nolimits_i\lambda_i^2 p_i^2|\phi_i|^{2p_i-2}}{\left(\sum\nolimits_{j} \lambda_j |\phi_j|^{p_j}\right)^2},\\
      r &=\frac{8}{\sum\nolimits_i\phi_{i}^2/p_i^2},\label{r}\\
      \frac{n_t}{r}&=-\frac{1}{8}\frac{\sum\nolimits_i\lambda_i^2 p_i^2|\phi_i|^{2p_i-2}}{\left(\sum\nolimits_{j} \lambda_j |\phi_j|^{p_j}\right)^2}\sum\nolimits_l\frac{\phi_{l}^2}{p_l^2}.\label{consistency}
\end{align}


We set up the probability distribution for the parameters Eq. \eqref{ns}-\eqref{consistency} by marginalizing them over $P(\lambda)$, $P(\phi_*)$, and $P(p)$ and then calculate their expectations and corresponding variances by applying the central limit theorem (CLT) in many-field limit in the order of magnitude about $N_f>\mathcal{O}(100)$. To further simplify the expressions, we boost $N_f$ to be really large in the order $\mathcal{O}(10^{40})$ and use Laplace method to produce the final analytical results precisely. Different choice of  initial conditions has an insignificant effect on the density spectra \cite{Easther:2013rva}. And applying the HCA in Eq. \eqref{NtoField} implies that P$(\phi_*)$ is a uniform prior on the surface of an $N_f$ ellipsoid whose elliptic radii are determined by P$(p)$. So we can sample the ellipsoid by defining
\begin{align}
\phi_{i}=\sqrt{\frac{2p_iN_*}{\sum\nolimits_j x_j^2}}x_i\quad \text{for}\quad \vec{x}\sim {\mathcal{N}}(0,\mathbbm{1}),
\end{align}
where ${\mathcal{N}}(0,\mathbbm{1})$ is a multivariate normal distribution. Subsequently, one of the summations in Eq. \eqref{ns}-\eqref{consistency} is
\begin{align}\label{summation1}
\sum\nolimits_i\lambda_i^2 p_i^2|\phi_i|^{2p_i-2}=\sum_i\frac{\lambda_i^22^{p_i-1}p_i^{p_i+1}N_*^{p_i-1}|x_i|^{2p_i-2}}{\sqrt{\sum\nolimits_j x_j^2}^{(2p_i-2)}}.
\end{align}
In many-field limit $N_f\to\infty$ the CLT ensures that the summation is normally distributed with mean
\begin{align}
\left\langle\sum\nolimits_i\lambda_i^2 p_i^2|\phi_i|^{2p_i-2}\right\rangle_{N_f\uparrow}&=N_f\left\langle\lambda^2\right\rangle\left\langle\frac{2^{p-1}p^{p+1}N_*^{p-1}|x|^{2p-2}}{\sqrt{N_f}^{(2p-2)}}\right\rangle
\end{align}
in which we assume that $\lambda_i$, $p_i$ and $x_i$ are independent and angle brackets $\langle .\rangle$ indicates the expectation value. The lower term of denominator in Eq.  \eqref{summation1} $\sqrt{\sum\nolimits_j x_j^2}$ is $\chi$-distribution and approaches normal distribution $\mathcal{N}(\sqrt{N_f},1/\sqrt{2})$ in many-field limit.
Besides, for any normally distributed variable $x\sim \mathcal{N}(\mu,\sigma)$ \cite{momNormalRVs}
\begin{align}\label{normallyRVs}
\langle |x|^\nu\rangle=\frac{2^{(\nu/2)}\sigma^\nu}{\sqrt{\pi}}\Gamma\left(\frac{1+\nu}{2}\right)F_{1,1}\left(-\frac{\nu}{2};\half;-\frac{\mu^2}{2\sigma^2}\right),
\end{align}
where $F_{1,1}$ is the confluent hypergeometric function of the first kind and $\nu>-1$. If $\nu<-1$, $\langle |x|^\nu\rangle$ may diverge.
As for $\mu=0,~\sigma=1$, then $F_{1,1}=1$.

Also we know the ratio distribution $\alpha/\beta$ for normally distributed random variables (RVs) $\alpha\sim\mathcal{N}(\mu_\alpha,\sigma_\alpha)$ and $\beta\sim\mathcal{N}(\mu_\beta,\sigma_\beta)$ as $P(\beta>0)\to 1$ will approach a normal distribution with mean $\mu_{\alpha}/\mu_{\beta}$ and standard deviation \cite{NormalRVsRatio}
\begin{align}\label{deviationRatio}
s=\frac{\sqrt{\mu_{\beta}^2\sigma^2_{\alpha}-2\gamma\mu_{\alpha}\mu_{\beta}\sigma_{\alpha}\sigma_{\beta}+\mu_{\alpha}^2\sigma_\beta^2}}{\mu_\beta^2}
\end{align}
in many-field limit, where $\gamma\equiv\langle(\alpha-\mu_\alpha)(\beta-\mu_\beta)\rangle/(\sigma_\alpha\sigma_\beta)\in[-1,1]$ is the correlation. The term $(\sum_i\lambda_i|\phi_{i,*}|^{p})^2$ is also approximately normal in many-field limit and we can prove the relation
$
\left\langle\(\sum\nolimits_{i}\lambda_i|\phi_{i}|^{p_i}\)^2\right\rangle=\left\langle\sum\nolimits_i\lambda_i|\phi_{i}|^{p_i}\right\rangle^2$ as $N_f\to \infty$.
Then in many-field limit the mean of the summation 
\begin{align}
 \left\langle\sum\nolimits_i\lambda_i^2 p_i^2|\phi_i|^{2p_i-2}\right\rangle=\frac{N_f}{\sqrt{\pi}}\left\langle\lambda^2\right\rangle\left\langle \frac{N^{p-1}_*}{N^{p-1}_f}2^{2p-2}p^{p+1}\Gamma(p-\half)\right\rangle\label{summation}
\end{align}
is finite when $p>1/2$. The means of other summations are similar to Eq. \eqref{summation} and all the standard deviations can be calculated from the mean values and the corresponding two-moments so they are tedious algebraic functions of $\langle \lambda\rangle$, $\langle \lambda^2\rangle$, $\langle \lambda^4\rangle$ and many other terms.

Finally by applying Eq. \eqref{deviationRatio} and other conclusions in many-field limit the value of $r$ is normally distributed with a mean
\begin{align}\label{rResult}
       \langle r\rangle=\frac{4}{N_*\langle1/p\rangle},
\end{align}
and a standard deviation proportional to
\begin{align}
s_{r}&=\frac{1}{\sqrt{N_f}}\frac{4}{N_*}\frac{\sqrt{3\sigma_{1/p}^2+4\mu_{1/p}^2-2\sqrt{2}\gamma'\mu_{1/p}\sqrt{3\sigma_{1/p}^2+2\mu_{1/p}^2}}}{\mu_{1/p}^2}\nonumber\\ 
    &\propto \frac{1}{\sqrt{N_f}}\to 0 \quad \text{as}\quad N_f\to \infty,
\end{align}
where $\gamma'$ is the correlation between the numerator and denominator in Eq. \eqref{r}.
The value of $n_t$ is normally distributed with a mean
\begin{align}
        \langle n_t\rangle=-2\langle\epsilon\rangle&=-\frac{\sqrt{\pi}}{4N_*}\frac{\langle\lambda^2\rangle}{\langle\lambda\rangle^2}\frac{\left\langle2^{2p}(N_*/N_f)^{p}p^{p+1}\Gamma(p-\half)\right\rangle}{\left\langle2^{p}(N_*/N_f)^{(p/2)}p^{p/2}\Gamma(\frac{p}{2}+\half)\right\rangle^2}\label{ntNfExact}\\
    &\approx-\frac{\sqrt{\pi}}{4N_*}\frac{\langle\lambda^2\rangle}{\langle\lambda\rangle^2} \frac{p_{m}\Gamma(p_{m}-\frac{1}{2})}{4\Gamma^2\left(\frac{p_{m}+1}{2}\right)}\frac{1}{f(p_{m})}\ln \left(\frac{N_f}{N_*}\right),\label{ntNf}
\end{align}
where  $p_{m}$ is the minimum possible value and $f(p)$ is the probability density function (PDF) of $p$. Note that a finite prediction for mean requires $p>1/2$ and a finite standard deviation requires $p>3/4$. To get the approximation Eq. \eqref{ntNf} we have employed Laplace method (see Appendix \ref{laplaceMethod}). Both the standard deviations of $n_t$ and $\epsilon$ are proportional to
\begin{align}
s_{n_t}=4s_{\epsilon}\propto \frac{1}{\sqrt{N_f}}\to 0 \quad \text{as}\quad N_f\to \infty.
\end{align}

The value of consistency relation $n_t/r$ is a multiplication of two normally distributed asymptotic-sharp random variate with a mean
\begin{align}
\left\langle \frac{n_t}{r}\right\rangle_{N_f\uparrow}
   &=-\frac{1}{8}\frac{\langle\lambda^2\rangle}{\langle\lambda\rangle^2}\left\langle\frac{1}{p}\right\rangle\frac{\sqrt{\pi}}{2}\frac{\left\langle2^{2p}(N_*/N_f)^{p}p^{p+1}\Gamma(p-\half)\right\rangle}{\left\langle2^{p}(N_*/N_f)^{\frac{p}{2}}p^{p/2}\Gamma(\frac{p}{2}+\half)\right\rangle^2}\label{FinalResultExact}\\
   &\approx-\frac{1}{8}\frac{\langle\lambda^2\rangle}{\langle\lambda\rangle^2}\left\langle\frac{1}{p}\right\rangle \frac{\sqrt{\pi}p_{m}\Gamma(p_{m}-\frac{1}{2})}{8\Gamma^2\left(\frac{p_{m}+1}{2}\right)}\frac{1}{f(p_{m})}\ln \left(\frac{N_f}{N_*}\right),\label{FinalResult}
\end{align}
where the requirements for $p$ are the same as above and please see Eq. \eqref{splitProof} in Appendix \ref{sLargeNf} for the validity of multiplication splitting. Concretely, for typical $P(\lambda)$ and $P(p)$, Eq. \eqref{FinalResultExact} will be a very good approximation when $N_f$ is larger than $\mathcal{O}(100)$ but the approximation Eq. \eqref{FinalResult} is as good as Eq. \eqref{FinalResultExact} generally only if $N_f$ is larger than $\mathcal{O}(e^{100})\sim\mathcal{O}(10^{40})$. The standard deviation is proportional to
\begin{align}
s_{n_t/r}\propto \frac{1}{\sqrt{N_f}}\to 0 \quad \text{as}\quad N_f\to \infty,\label{sofnt_r}
\end{align}
and also see Appendix \ref{sLargeNf} for the detailed proof.

The value of $n_s$ is a combination of two normally distributed variate with a mean
\begin{align}
 \langle n_s\rangle-1=\frac{-1}{N_*\langle1/p\rangle}\frac{\left\langle(N_*/N_f)^{p/2} 2^{p}p^{(p/2-1)}\Gamma(\frac{p}{2}+\half)\right\rangle}{\left\langle(N_*/N_f)^{p/2} 2^{p}p^{p/2}\Gamma(\frac{p}{2}+\half)\right\rangle}-2\langle\epsilon\rangle\label{nsFinalExact}\\
       \approx-\frac{1}{N_*\langle1/p\rangle p_{m}}-\frac{\sqrt{\pi}}{4N_*}\frac{\langle\lambda^2\rangle}{\langle\lambda\rangle^2} \frac{p_{m}\Gamma(p_{m}-\frac{1}{2})}{4\Gamma^2\left(\frac{p_{m}+1}{2}\right)}\frac{1}{f(p_{m})}\ln \left(\frac{N_f}{N_*}\right).\label{FinalScalarIndex}
\end{align}
and the standard deviation of the left term of the result is also proportional to
\begin{align}
s_{L}\propto \frac{1}{\sqrt{N_f}}\to 0 \quad \text{as}\quad N_f\to \infty.
\end{align}

From Eq. \eqref{ntNf}, requiring the slow variation parameter $\epsilon\lesssim 0.1$ then sets the upper limit of $N_f$
\begin{align}
N_f\lesssim N_*\exp\(ZN_*\)
\end{align}
where $Z$ is a value depends on the specific probability distributions of $\lambda_i$ and $p_i$
\begin{align}
Z=\frac{8}{\sqrt{\pi}}\frac{\langle\lambda\rangle^2}{\langle\lambda^2\rangle}\frac{4\Gamma^2\(\frac{p_m+1}{2}\)}{p_m\Gamma\(p_m-\frac{1}{2}\)}f(p_m)\times\mathcal{O}\(10^{-1}\).
\end{align}
 In addition, combining  Eq. \eqref{rResult}, Eq. \eqref{ntNf}, Eq. \eqref{FinalResult} and Eq. \eqref{FinalScalarIndex} immediately reaches the lower limiting value of consistency relation $n_t/r$
\begin{align}
\left\langle \frac{n_t}{r}\right\rangle\gtrsim -\frac{N_*}{2}\left\langle\frac{1}{p}\right\rangle\times\mathcal{O}(10^{-1}),
\end{align}
and a relation
\begin{align}
\langle n_s\rangle=1-\frac{\langle r\rangle}{4p_m}+
\langle n_t\rangle
\end{align}
 which is independent of specific probability distribution of $\lambda_i$, $p_i$ and $\phi_{i,*}$. Adding the restriction $p_m>1/2$ gives two bounds of $r$ 
 \begin{align}
 r&>\frac{2}{N_*},\label{rRestriction}\\
 r&>2(1-n_s+n_t),
 \end{align}
 and the value range of $n_t$ as
 \begin{align}
 \frac{1}{2}\frac{1}{p_mN_*}+n_s-1<n_t<0
 \end{align}
 which can be tested by observation in the near future because Eq. \eqref{rRestriction} indicates $r\gtrsim0.03$, which is exactly on the coverage of the next generation projects under construction. 

 \begin{figure}
  \includegraphics[scale=0.5]{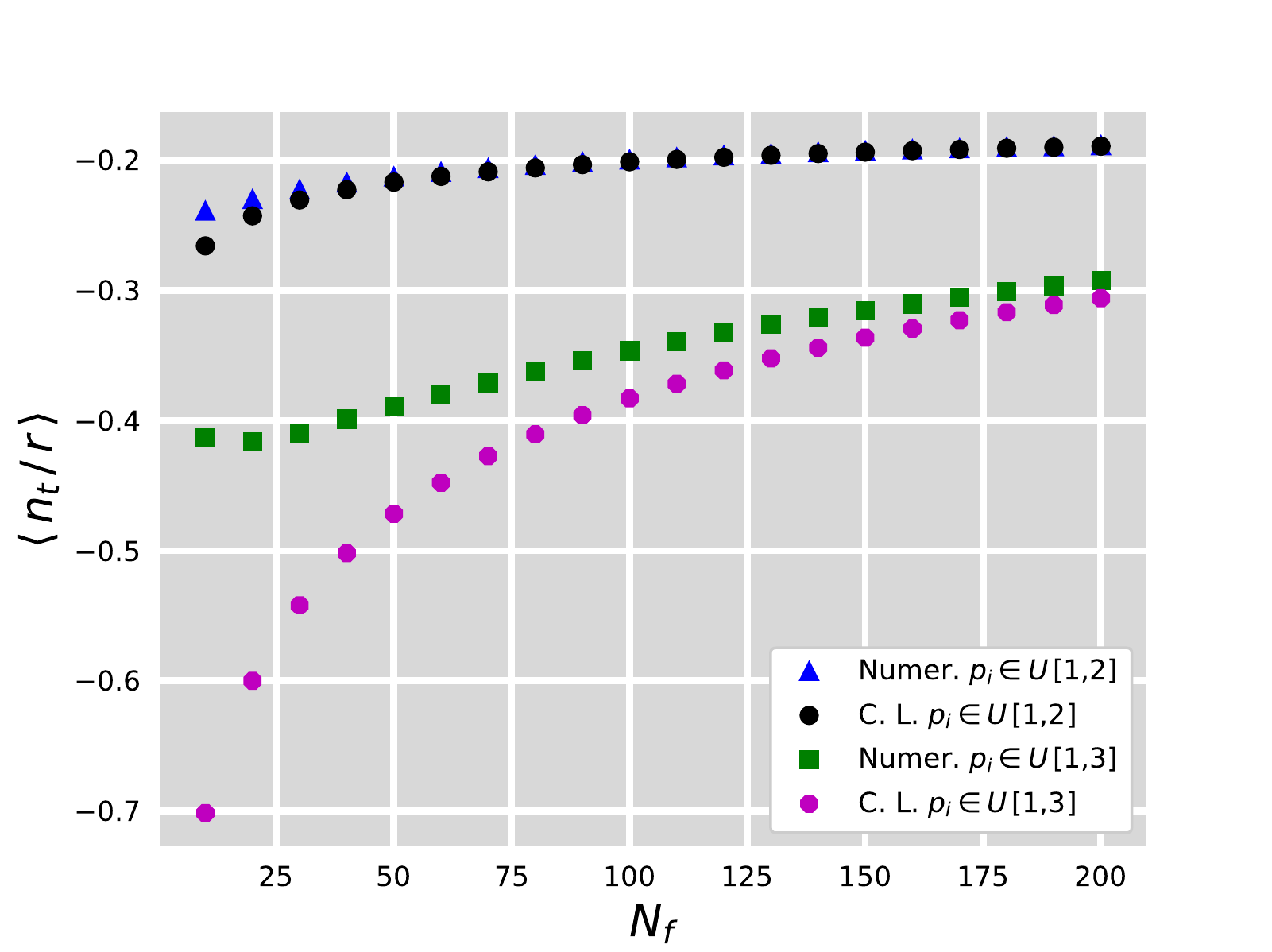}
  \caption{The multifield prediction from CLT in Eq. \eqref{FinalResultExact} compared to the numerical simulations with 100\,000 samples, $\lambda_i\in\mathcal{U}[10^{-14},10^{-13}]$, $p_i\in\mathcal{U}[1,2]$ and $p_i\in\mathcal{U}[1,3]$. Using the horizon-crossing approximation, the field values $\phi_{i,*}$ as the pivot scale $k_*$ leaves the horizon are originated from a uniform prior on the surface in Eq. \eqref{NtoField}.}
  \label{fig1}
\end{figure}
Obviously, with all $p_i$ equal we can regain all the conclusions described in \cite{Price:2014ufa} and many other classic results from Eq. \eqref{FinalResultExact}, Eq. \eqref{rResult}, Eq. \eqref{ntNfExact},  and Eq. \eqref{nsFinalExact}. But the extent of deviation from the single-field model result of $n_t/r=-1/8$ gets much larger than the fixed-$p$ ones.  Figure \ref{fig1} compares the predicted value from CLT for $\langle n_t/r\rangle$ in Eq. \eqref{FinalResultExact} to corresponding numerical results from Eq. \eqref{consistency} with uniform-distribution $\lambda_i\in\mathcal{U}[10^{-14},10^{-13}]$ and uniform-distribution $p_i\in\mathcal{U}[1,2]$ and $p_i\in\mathcal{U}[1,3]$ respectively, showing  excellent convergence in many-field limit. Furthermore, the wider the distribution of $p_i$ is, the larger $N_f$ is needed for getting the comparable convergence. Also we can strictly prove that the corresponding relative error is proportional  to $1/N_f$.
 \begin{figure}[htbp]
\includegraphics[scale=0.5]{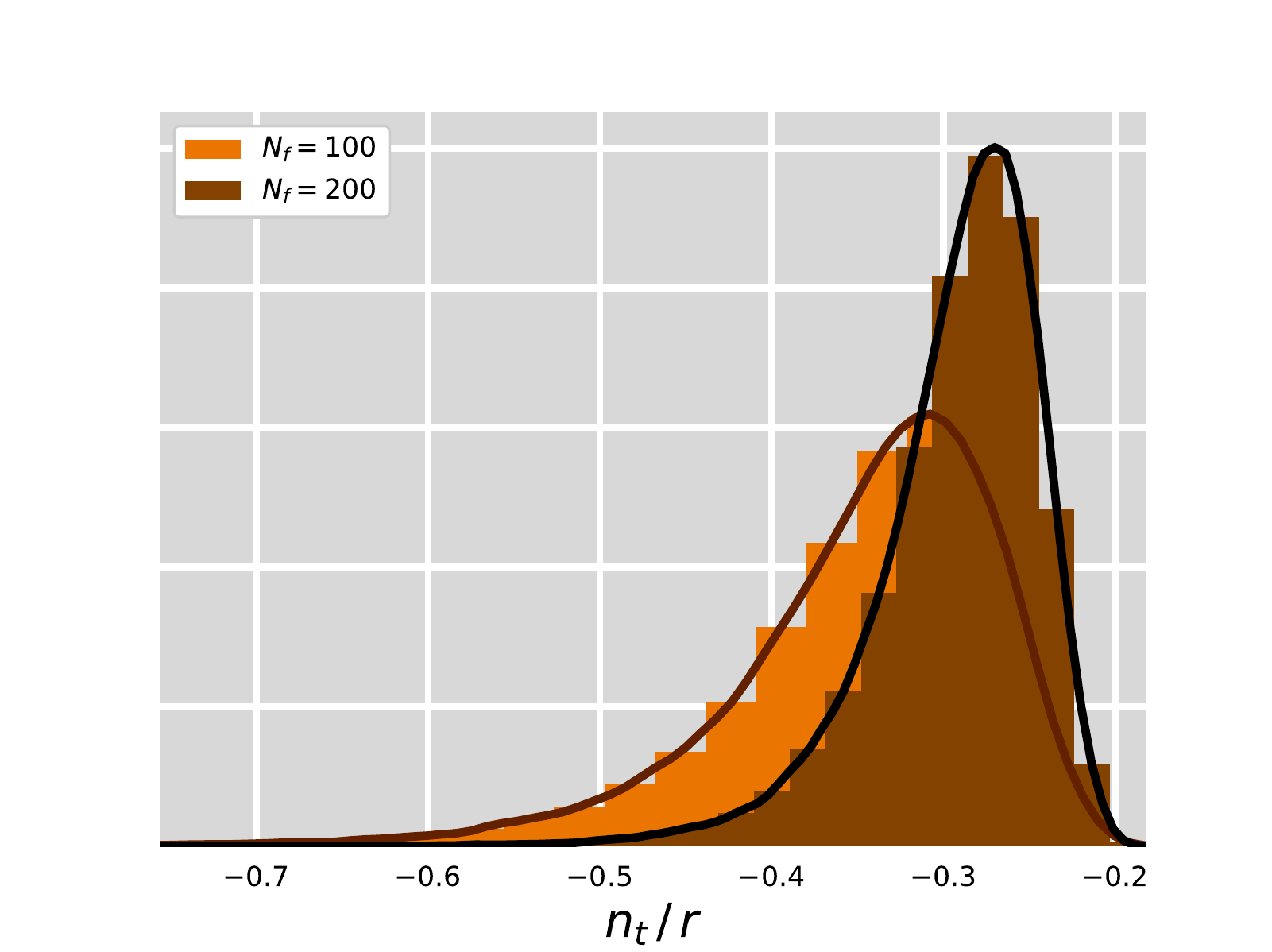}
\caption{The probability distributions for $n_t/r$ with histograms built from 100\,000 numerical samples with $\lambda_i\in\mathcal{U}[10^{-14},10^{-13}]$ and $p_i\in\mathcal{U}[1,3]$ when $N_f=100$ and $N_f=200$.   }\label{fig2}
\end{figure}

 Figure \ref{fig2} delineates the PDF for $n_t/r$ with  $\lambda_i\in\mathcal{U}[10^{-14},10^{-13}]$ and $p_i\in\mathcal{U}[1,3]$ when $N_f$ is 100 and 200 respectively. As shown, the larger the $N_f$ becomes, the sharper the PDF of $n_t/r$ will be and the more likely that the mean of $n_t/r$ can well represent the real value, as proved in  Eq. \ref{sofnt_r}. 
 \begin{figure}[htbp]
\includegraphics[scale=0.5]{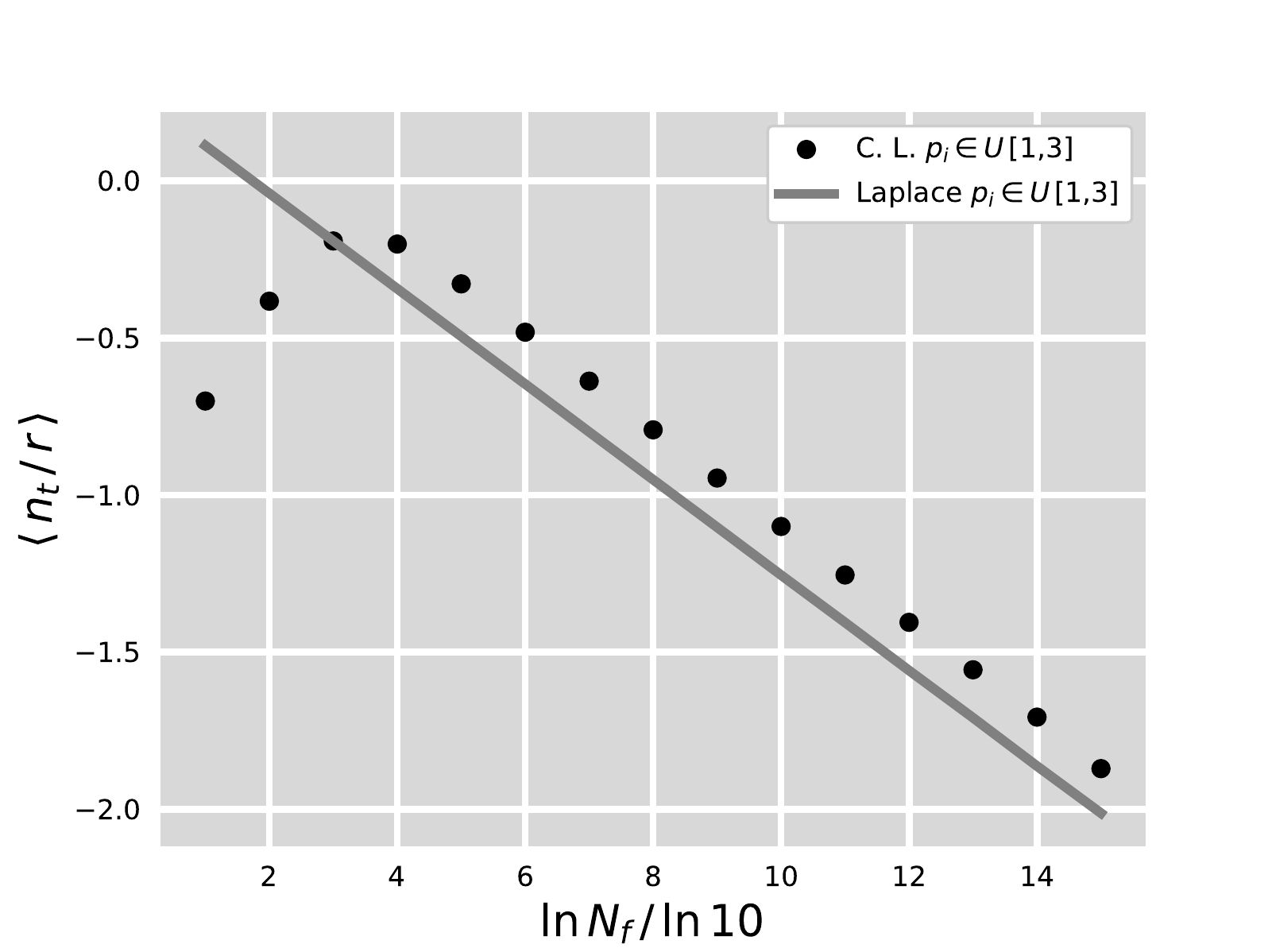}
\caption{The multifield analytic prediction from Laplace approximation method in Eq. \eqref{FinalResult} compared to the central limit results with 1\,000\,000 samples, $\lambda_i\in\mathcal{U}[10^{-14},10^{-13}]$ and $p_i\in\mathcal{U}[1,3]$. The logarithmic  correlation relation is evident in extremely large $N_f$.}\label{fig3}
\end{figure}

To understand the Laplace approximation result in Eq. \eqref{FinalResult} more intuitionally, we compare the central limit results for $\langle n_t/r\rangle$ in Eq. \eqref{FinalResultExact} to the predicted analytical values from Eq. \eqref{FinalResult} when $N_f$ is extremely large in Figure \ref{fig3}. It is clearly observed that when $N_f$ is small, the Laplace approximation is at a great deviation while the extremely large $N_f$ leads to a good agreement with Eq. \eqref{FinalResultExact} and the logarithmic  correlation relation is evident. Also a wider distribution of $p_i$ needs a larger $N_f$ for a good approximation. But in a narrow distribution of $p_i$ as the setup in the figure, $N_f$ needs not to be as large as $\mathcal{O}(10^{40})$ to make the Laplace approximation valid, only $N_f\gtrsim\mathcal{O}(10^{5})$. The $N_f\sim\mathcal{O}(10^{40})$ condition is suitable for general cases. Notice that the C. L. results, in such a large $N_f$, can represent the numerical ones perfectly well according to the aforementioned analysis.

\section{Conclusions}\label{conclusions}

We have computed the probability distributions for the tensor spectral index $n_t$, tensor-to-scalar ratio $r$, scalar spectral index $n_s$, and the consistency relation $n_t/r$ in the general large number monomial multifield inflation model, as a function of the probability distribution of couplings $\lambda_i$, power indexes $p_i$, initial field values and the number of fields $N_f$. In many-field limit, all the distributions become sharp with the variances $s^2\propto 1/N_f$, so the expected values we get are very robust.
 
 We give a novel prediction that the inflationary parameters $\epsilon$, $n_t$, $n_s$ and $n_t/r$  are all proportional to $\ln N_f$ when $N_f$ is extremely large. The dependency between $\epsilon$ and $\ln N_f$ immediately gives the upper bound of $N_f\lesssim N_*e^{ZN_*}$ if we require $\epsilon$ small enough such as $\mathcal{O}(10^{-1})$ where $Z$ is a value decided by the specific probability distributions of $\lambda_i$ and $p_i$. But the tensor-to-scalar ratio $r=4/\left(N_*\langle 1/p\rangle\right)$ depends only on the probability distribution of $p_i$.

 Besides, we find some distribution-independent relations between the inflationary observables and thereby we give some theoretical  bounds for $r$ and $n_t$ especially $r>2/N_*\ (r\gtrsim 0.03)$ which can be tested by observation in the near future. All predictions above together can distinguish diverse-$p$-$N_f$-monomial models, fixed-$p$-$N_f$-monomial models and their single-field analogues. This work marks another significant step in the multifield scenario where the predictions are sharp and generic in large-$N_f$ limit \cite{Price:2014ufa,Aazami:2005jf,Alabidi:2005qi,Piao:2006nm,Kim:2006te,Frazer:2013zoa,Kaiser:2013sna,Kallosh:2013daa,Wenren:2014cga,Sloan:2014jra,Price:2015qqb,Amin:2017wvc,Dias:2017gva}. Additionally, exploring a broader class of large number multifield models such as the multifield extension to small-field inflation will be intriguing follow-up work, in order to advance our understanding of the very early universe and the physics in extremely high energy.

\vspace{0.5 cm}
\noindent {\bf Acknowledgments}

The author would like to thank Qing-Guo Huang for careful review, comments, and feedback on this paper. The author is also grateful to Shi Pi and Cheng Cheng for helpful discussions. The contribution of HPC Cluster of ITP-CAS is highly appreciated. This work is supported by the project 11647601 of National Natural Science Foundation of China. 
\appendix
\section{Laplace method}
\label{laplaceMethod}
\setcounter{equation}{0}
We define the notations
\begin{gather*}
M\equiv \frac{N_f}{N_*}\to \infty,\\
g_1(p)\equiv 2^{2p}p^{p+1}\Gamma(p-\frac{1}{2})f(p),\quad
g_2(p)\equiv 2^{p}p^{p/2}\Gamma(\frac{p}{2}+\frac{1}{2})f(p),
\end{gather*}
where $f(p)$ is the PDF of $p$ and we restrict $p>1/2$. Thus we can rewrite one of the upper average terms in Eq. \eqref{ntNfExact} and Eq. \eqref{FinalResultExact} by 
\begin{align}
\left\langle2^{2p}(N_*/N_f)^{p}p^{p+1}\Gamma(p-\half)\right\rangle &=\int_{p_{m}}^{p_{max}}e^{-p\ln M}g_1(p)dp,\nonumber
\end{align}
where $p_{m}$ and $p_{max}$ are the minimum and the maximum possible value for typical possibility distribution of $p$, respectively. Then when $M\to\infty$ we use Laplace method to simplify the expression
\begin{align}
\int_{p_{m}}^{p_{max}}e^{-p\ln M}g_1(p)dp&\approx g_1(p_{m})\int_{p_{m}}^{p_{max}}e^{-p\ln M}dp\nonumber\\
    &\approx g_1(p_{m})e^{-p_{m}\ln M}/\ln M,\nonumber
\end{align}
where we have dropped the $p_{max}$ term because it decreases much faster than the $p_{m}$ term when $M\to\infty$
and the other average term is
\begin{align*}
\left\langle2^{p}(N_*/N_f)^{\frac{p}{2}}p^{p/2}\Gamma(\frac{p}{2}+\half)\right\rangle&\approx 2g_2(p_{min})e^{-\frac{p_{min}}{2}\ln M}/\ln M,
\end{align*}

Finally the $p$-average terms in Eq. \eqref{ntNfExact} and Eq. \eqref{FinalResultExact} can be expressed in large-$N_f$ limit by 
\begin{align*}
\frac{\left\langle2^{2p}(N_*/N_f)^{p}p^{p+1}\Gamma(p-\half)\right\rangle}{\left\langle2^{p}(N_*/N_f)^{\frac{p}{2}}p^{p/2}\Gamma(\frac{p}{2}+\half)\right\rangle^2}\Bigg|_{N_f\uparrow}&\approx \frac{g_1(p_{min})}{4g_2(p_{min})^2}\ln M\\
   =\frac{p_{min}\Gamma(p_{min}-\frac{1}{2})}{4\Gamma^2\left(\frac{p_{min}+1}{2}\right)}&\frac{1}{f(p_{min})}\ln \left(\frac{N_f}{N_*}\right).
\end{align*}

Similarly, the $p$-average term in Eq. \eqref{nsFinalExact} is
\begin{align*}
\frac{\left\langle(N_*/N_f)^{p/2} 2^{p}p^{(p/2-1)}\Gamma(\frac{p}{2}+\half)\right\rangle}{\left\langle(N_*/N_f)^{p/2} 2^{p}p^{p/2}\Gamma(\frac{p}{2}+\half)\right\rangle}\Bigg|_{N_f\uparrow}\approx \frac{1}{p_{min}}.
\end{align*}
\section{Standard deviation $s$ in large-$N_f$ limit}
\label{sLargeNf}
\setcounter{equation}{0}
Generally we suppose there are two normally distributed RVs $X_1\sim \mathcal{N}(\mu_1,\sigma_1)$, $X_2\sim \mathcal{N}(\mu_2,\sigma_2)$ and both standard deviations are inversely proportional to the square root of $N_f$ ($
\sigma_1\propto 1/\sqrt{N_f},\ \sigma_2\propto 1/\sqrt{N_f}$)
and the correlation coefficient of $X_1$ and $X_2$
$\gamma_{x_1x_2}=\text{Cov}(X_1,X_2)/\sigma_1\sigma_2<1,$
as a consequence of Cauchy-Bunyakovsky-Schwarz inequality. Immediately,
\begin{align}\label{splitProof}
\langle X_1X_2\rangle=\mu_1\mu_2+\gamma_{x_1x_2}\sigma_1\sigma_2\to \mu_1\mu_2\quad \text{as}\quad N_f\to \infty.
\end{align}
 Besides, the correlation coefficient of $X_1^2$ and $X_2^2$ is
$\gamma_{x_1^2x_2^2}=\text{Cov}(X_1^2,X_2^2)/\sigma_{X_1^2}\sigma_{X_2^2}$,
from which we can get
\begin{align}
\langle X_1^2X_2^2\rangle=\sigma_{X_1^2}\sigma_{X_2^2}\gamma_{x_1^2x_2^2}+(\sigma_1^2+\mu_1^2)(\sigma_2^2+\mu_2^2),\label{meanofSquareSquare}
\end{align}
and also  $|\gamma_{x_1^2x_2^2}|<1.$

Note from Eq. \eqref{normallyRVs} that for any normally distributed variable $X\sim \mathcal{N}(\mu,\sigma)$ we have $
\langle X^4\rangle=3\sigma^4+6\mu^2\sigma^2+\mu^4$, 
because $F_{1,1}(-2;\frac{1}{2};z)=1-4z+\frac{4}{3}z^2$.
Hence
\begin{align}
\sigma_{X_1^2}^2&=\langle X_1^4\rangle-\langle X_1^2\rangle^2=2\sigma_1^4+4\mu_1^2\sigma_1^2,\label{sigma1}\\
\sigma_{X_2^2}^2&=2\sigma_2^4+4\mu_2^2\sigma_2^2.\label{sigma2}
\end{align}
Then substitute Eq. \eqref{sigma1} and Eq. \eqref{sigma2} into Eq. \eqref{meanofSquareSquare} we get the standard deviation of the multiplication $X_1X_2$ 
\begin{align*}
s^2(X_1X_2)&=\langle X_1^2X_2^2\rangle-\langle X_1X_2\rangle^2\\
        =&\gamma_{x_1^2x_2^2}\(4\sigma_1^4\sigma_2^4+8\mu_1^2\sigma_1^2\sigma_2^4+8\mu_2^2\sigma
        _1^4\sigma_2^2+4\mu_1^2\mu_2^2\sigma_1^2\sigma_2^2\)^{1/2}\\
        &+(1-\gamma_{x_1x_2}^2)\sigma_1^2\sigma_2^2+\mu_1^2\sigma_2^2+\mu_2^2\sigma_1^2-2\gamma_{x_1x_2}\mu_1\mu_2\sigma_1\sigma_2\\
        \propto& \frac{1}{N_f} \quad \text{as}\quad N_f\to \infty,
\end{align*}
if both $\mu_1$ and $\mu_2$ are not go up faster than $N_f^q$ where $q$ is an arbitrary positive number.

For the term in Eq. \eqref{ns}
\begin{align*}
X_1=\frac{1}{\sum\nolimits_i\phi_{i,*}^2/p_i^2},\quad 
X_2=\frac{\sum\nolimits_j(\lambda_j/p_j)|\phi_j|^{p_j}}{\sum\nolimits_{l} \lambda_l |\phi_l|^{p_l}}.
\end{align*}
Then in large-$N_f$ limit
\begin{align*}
\mu_1\to \frac{1}{2N_*\langle 1/p\rangle}\propto N_f^0<N_f^q,\quad \mu_2 \to \frac{1}{p_{min}}\propto N_f^0<N_f^q.
\end{align*}

For the term in Eq. \eqref{consistency}
\begin{align*}
X_1=\frac{\sum\nolimits_i\lambda_i^2 p_i^2|\phi_i|^{2p_i-2}}{\left(\sum\nolimits_{j} \lambda_j |\phi_j|^{p_j}\right)^2},\quad 
X_2=\sum\nolimits_l\frac{\phi_{l,*}^2}{p_l^2}.
\end{align*}
In large-$N_f$ limit
\begin{align*}
\mu_1\propto \ln N_f<N_f^q,\quad \mu_2 \to Const.\propto N_f^0<N_f^q.
\end{align*}
Obviously we have proved the asymptotic inverse square root relation
\begin{align*}
s_{L},\  s_{n_t/r}\propto \frac{1}{\sqrt{N_f}},\quad \text{as}\quad N_f\to \infty.
\end{align*}

\newpage


\end{document}